\documentclass[prd,superscriptaddress,reprint,aps,amsmath,amssymb,nofootinbib]{revtex4-2}  % 删除了 showpacs

\usepackage{graphicx}
\usepackage{enumerate, xcolor, enumitem, xspace, braket}
\usepackage{booktabs}
\usepackage{makecell}
\usepackage{multirow}
\usepackage{comment}
\usepackage{rotating}
\usepackage[normalem]{ulem}
\usepackage{tikz-feynman}
\usepackage[colorlinks=true,citecolor=blue]{hyperref}

\begin{document}

\title{Revealing the Two-Fold Ambiguity: Tau Momentum Reconstruction and Its Impact on Entanglement Observables
}

\author{Xiang Zhou}
\email{20241792@snnu.edu.cn}
\affiliation{School of Physics $\&$ Information Technology,
Shaanxi Normal University, Xi’an 710119, China}

\author{Jianyong Zhang}
\email{jyzhang@ihep.ac.cn}
\affiliation{ Institute of High Energy Physics, 
\\Chinese Academy of Sciences, Beijing 100049, China} 

\author{Xia Wan}
\email{wanxia@snnu.edu.cn}
\affiliation{School of Physics $\&$ Information Technology,
Shaanxi Normal University, Xi’an 710119, China}

\author{Youkai Wang}
\email{wangyk@snnu.edu.cn}
\affiliation{School of Physics $\&$ Information Technology,
Shaanxi Normal University, Xi’an 710119, China}

\author{Xiaohu Mo}
\email{moxh@ihep.ac.cn}
\affiliation{ Institute of High Energy Physics, 
\\Chinese Academy of Sciences, Beijing 100049, China} 
\affiliation{ University of Chinese Academy of Science, Beijing 100049, China} 
\date{\today}
%{\hfill PITT-PACC-2412}

\begin{abstract}
The neutrinos produced in $\tau$ decays cannot be directly detected, making the reconstruction of $\tau$ kinematics challenging and affecting measurements of quantum correlations such as spin entanglement. For the process $e^+e^- \to \tau^+\tau^- \to \pi^+ \bar{\nu}_\tau\pi^-\nu_\tau$, the kinematic constraints allow the $\tau$ momenta to be reconstructed up to a well-known two-fold ambiguity, regardless of the presence of an intermediate resonance state. In this paper, we present a geometric interpretation of this ambiguity and propose a numerical reconstruction method based on singular value decomposition (SVD). Using only the information from visible final-state particles and decay kinematics, the method reconstructs the two possible solutions for the $\tau^+\tau^-$ pair. The reconstruction performance is validated with Monte Carlo simulations in typical collider environments. We further investigate the impact of the spurious solution on spin-entanglement measurements and show that reliable entanglement signals can still be extracted even when the true and spurious solutions cannot be experimentally distinguished. This work provides a practical approach for $\tau$-lepton kinematic reconstruction and spin-entanglement measurements in $e^+e^-$ collider experiments.

%The neutrinos produced in $\tau$ lepton decays cannot be directly detected, making it impossible to reconstruct the kinematic information of the $\tau$ lepton, which severely affects the measurement of quantum correlations such as spin entanglement. In this paper, we first clearly illustrate, from geometric and numerical perspectives, the reconstruction process of the $\tau$ momentum and the  origin of the “two-fold ambiguity” problem. On this basis, we propose a numerical method based on singular value decomposition (SVD) that uses the kinematic constraints of the decay to solve for the $\tau$ momentum from the information of visible final-state particles. The reconstruction accuracy of the method is verified through Monte Carlo simulations in typical collider environments. Furthermore, we apply the proposed method to the measurement of spin entanglement and discuss in detail the impact of the spurious solution in the two-fold ambiguity on the entanglement signal. The results show that even if the true and spurious solutions cannot be distinguished experimentally, the spin entanglement measurement can still be reliably performed. This study provides a feasible approach for conducting $\tau$ lepton spin entanglement measurements when the two-fold ambiguity cannot be resolved, and also offers an automated means of reconstructing the $\tau$ kinematic information for entanglement measurement experiments.
\end{abstract}

\maketitle
\newpage
{
\hypersetup{linkcolor=blue}
%\tableofcontents
}

%%%%%%%%%%%%%%%%%%%%%%%%%%%%%%%%%%%%%%%%%%%%%%%%%%%%%%%%%%%
%%%%%%%%%%%%%%%%%%%%%%%%%%%%%%%%%%%%%%%%%%%%%%%%%%%%%%%%%%%
\section{Introduction}
Since 2020, measurements of particle-pair entanglement~\cite{Afik2021} and Bell nonlocality ~\cite{Fabbrichesi:2021npl,Severi:2021cnj,Aoude:2022imd,Afik:2022kwm,Aguilar-Saavedra:2022uye,Fabbrichesi:2022ovb,Severi:2022qjy,Dong:2023xiw,Han:2023fci} in collider experiments have become increasingly active. The ATLAS~\cite{ATLAS:2023fsd} and CMS~\cite{CMS:2024pts} collaborations at the LHC have successively observed spin entanglement in top-quark pairs, while the BESIII collaboration has measured hadron-pair entanglement in charmonium decays~\cite{BESIII:2025vsr}. These achievements have verified the predictions of quantum mechanics regarding entanglement in the quark and baryon sectors, further motivating expectations for similar quantum property tests in the lepton sector.

Among leptons, the $\tau$ lepton~\cite{Privitera:1992ja, Abel:1992kz, Dreiner:1992gt, Fabbrichesi:2024wcd} is undoubtedly an ideal object for studying spin entanglement. However, a key challenge in measuring lepton entanglement is the reconstruction of its momentum: since $\tau$ decay products include neutrinos that cannot be directly detected in the detector, the kinematic state of the $\tau$ must be inferred solely from the information of visible final-state particles. %%(such as charged hadrons, $\pi$ mesons). 
Moreover, previous studies~\cite{Kuhn:1993ra} often encountered a “two-fold ambiguity" in the reconstruction of $\tau$ momentum. However, the existing expressions~\cite{Fabbrichesi:2024wcd,Kuhn:1993ra} are not conducive to intuitively identifying this two-fold ambiguity, nor are they convenient for experimentalists to implement automated reconstruction.

To address this, we clearly explain the origin of the two-fold ambiguity from geometric and numerical perspectives, and propose a numerical method based on singular value decomposition (SVD) to solve the corresponding non-square system of equations, thereby reconstructing the $\tau$ momentum. This method not only clearly reveals the mathematical origin of the two-fold ambiguity but also verifies the high precision of the obtained solutions through Monte Carlo simulations. In addition, we discuss the impact of the spurious solution in the two-fold ambiguity on spin entanglement, and find that it does not affect the measurement of spin entanglement. This means that even if the two solutions cannot be experimentally distinguished at current colliders, spin-entanglement measurements of the $\tau$ pair can still be reliably performed. The proposed approach therefore provides a robust momentum-reconstruction tool for experimental studies of $\tau$-pair spin entanglement.

The structure of this paper is as follows: Section~\ref{sec:2} illustrates the two-fold ambiguity in $\tau$-momentum reconstruction from geometric and numerical perspectives, and presents the detailed framework of the SVD-based reconstruction method; Section~\ref{sec:3} verifies the performance of the method in terms of momentum reconstruction accuracy via Monte Carlo simulations; Section~\ref{sec:4} applies the method to spin-entanglement measurements, discusses the impact of the spurious solution on the entanglement signal, and shows that reliable entanglement measurements can still be achieved when both solutions of the two-fold ambiguity are retained.

%%%%%%%%%%%%%%%%%%%%%%%%%%%%%%%%%%%%%%%%%%%%%%%%%%%%%%%%%%%
%%%%%%%%%%%%%%%%%%%%%%%%%%%%%%%%%%%%%%%%%%%%%%%%%%%%%%%%%%%
\section{Methodology}\label{sec:2}
%BESIII possesses the world's largest data sample of $\tau$-charm physics~\cite{BESIII:2020nme, BESIII:2022mxl, BESIII:2014srs}, so this paper primarily takes its energy region as an example.

BESIII possesses the world’s largest data sample in $\tau$-charm physics~\cite{BESIII:2020nme, BESIII:2022mxl, BESIII:2014srs}, providing an excellent opportunity for experimental studies of $\tau$-pair entanglement. Motivated by this experimental advantage, we illustrate the kinematic framework for $\tau$-momentum reconstruction using the representative process
\begin{equation}
e^+e^- \to \psi(2S) \to \tau^+\tau^- \to \pi^+\bar{\nu}_\tau \pi^-\nu_\tau ,
\end{equation}
while the reconstruction method itself is applicable to generic $e^+e^- \to \tau^+\tau^-$ processes with the same final states, provided that the initial $e^+e^-$ four-momentum is known.

In the rest frame of $\psi(2S)$, the energies of $\tau^+$ and $\tau^-$ are equal, and their momenta have equal magnitude but opposite directions:
\begin{equation}
E_{\tau^+}=E_{\tau^-}=\frac{m_{\psi(2S)}}{2},
\end{equation}
\begin{equation}
|\vec{p}_{\tau^+}|=|\vec{p}_{\tau^-}|=\sqrt{\left(\frac{m_{\psi(2S)}}{2}\right)^2-m_\tau^2}.
\end{equation}

\subsection{Kinematic Constraint of a Single $\tau$ Decay}
\begin{figure}
  \centering
  \includegraphics[width=0.45\linewidth]{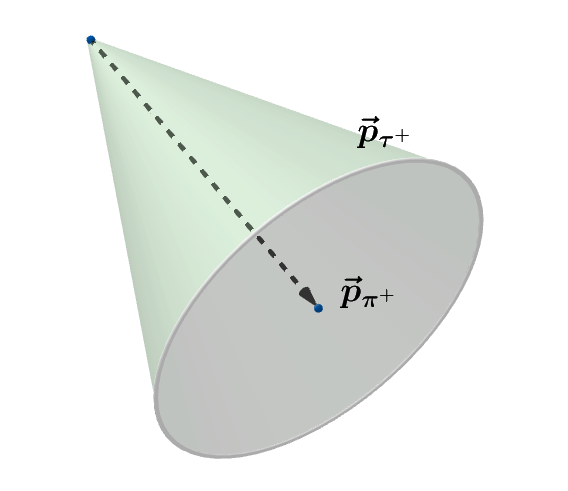}
  \qquad
  \includegraphics[width=0.43\linewidth]{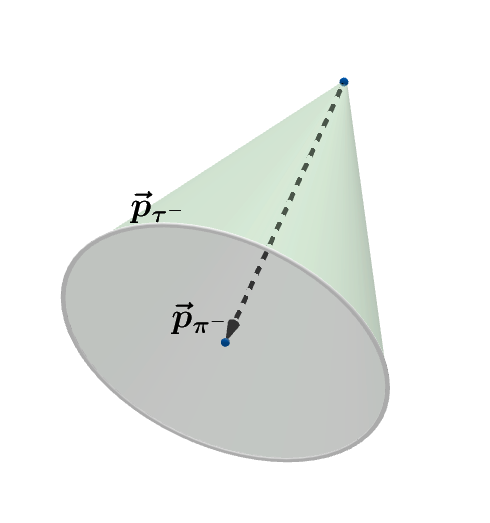}
  \caption{Illustration of the constraint cones for the $\tau$ momentum direction.  $\tau^+$ constraint cone (left); $\tau^-$ constraint cone (right). The cones represent all possible directions of the $\tau$ momentum, while the black arrowed rays indicate the momentum directions of the corresponding pions.}
  \label{fig:cone}
\end{figure}

Taking $\tau^{+}\rightarrow\pi^{+}\bar{\nu}_{\tau}$ as an example, four-momentum conservation leads to 
\begin{equation}
p_{\tau^+}-p_{\pi^+}=p_{\bar\nu_\tau}~,
\end{equation}
squaring both sides, and noticing the zero mass of $\tau$ neutrino ($p_{\bar\nu_{\tau}}^{2}=0$), it yields
\begin{equation}
p_{\tau^+}\cdot p_{\pi^+}=\frac{m_\tau^2+m_\pi^2}{2}~,
\end{equation}
which implies that in the rest frame, the dot product between the $\tau^+$ momentum and the $\pi^+$ momentum is constant. Rewriting the above equation as: 
\begin{equation}
E_{\tau^+}E_{\pi^+}-|\vec{p}_{\tau^+}||\vec{p}_{\pi^+}|\cos\theta_{\tau^+,\pi^+}=\frac{m_\tau^2+m_\pi^2}{2}~.\label{eq:5}
\end{equation}
Where $\theta_{\tau^+,\pi^+}$ denotes the angle between the momentum directions of $\tau^+$ and $\pi^+$. From Eq.~(\ref{eq:5}), given the momentum of $\pi^+$, the direction of the $\tau^+$ momentum must lie on a cone whose central axis is the $\pi^+$ momentum direction. The opening angle of the cone is determined by the known $E_{\tau^+}$, $|\vec{p}_{\tau^+}|$, and the energy and momentum of $\pi^+$. This effect is illustrated in Fig.~\ref{fig:cone}, where the ray represents the momentum direction of $\pi^+$, and the conical surface represents all possible $\tau^+$ momentum directions.

For the decay $\tau^-\to\pi^-\nu_\tau$, a similar process occurs: the direction of the $\tau^-$ momentum also lies on a cone whose axis is the $\pi^-$ momentum direction.

\subsection{$\tau^+ \tau^-$ Correlation and Geometric Interpretation of the Two-fold Ambiguity}
For the process $\psi(2S)$ decaying into a $\tau^+\tau^-$ final state, the two $\tau$ leptons are produced back-to-back in the rest frame of their parent particle, so we have
\begin{equation}
\vec{p}_{\tau^+}+\vec{p}_{\tau^-}=0,
\end{equation}
hence
\begin{equation}
\vec{p}_{\tau^-}\cdot\vec{p}_{\pi^-}=E_{\tau^-}E_{\pi^-}-\frac{m_\tau^2+m_\pi^2}{2}~.
\end{equation}
The angle between the $\tau^-$ momentum and the $\pi^-$ momentum can be transformed into the angle between the $\tau^+$ momentum and the opposite direction of the $\pi^-$ momentum:
\begin{equation}
-\vec{p}_{\tau^+}\cdot\vec{p}_{\pi^-}=E_{\tau^-}E_{\pi^-}-\frac{m_\tau^2+m_\pi^2}{2}.
\end{equation}
Thus, for $\tau^+$, both cone constraints must be satisfied simultaneously. The condition for the two cones to intersect is governed by their half-opening angles $\theta
_{\tau^+,\pi^+}$ and $\theta_{\tau^+,-\pi^-}$ and the angle $\Omega$ between their axes, as expressed by the inequality
\begin{equation}  
|\theta_{\tau^+,\pi^+}- \theta_{\tau^+,-\pi^-}| \le \Omega \le \theta_{\tau^+,\pi^+} + \theta_{\tau^+,-\pi^-}.\label{eq:iq}
\end{equation}
When $\Omega$ lies strictly between the bounds, the two cones intersect, yielding two possible momentum directions for $\tau^+$ . When $\Omega$ equals either bound, the cones are exactly tangent, giving a unique solution. When $\Omega$ lies outside this range, the cones do not intersect, and no physical solution exists.

\begin{figure}
  \centering
  \includegraphics[width=0.5\linewidth]{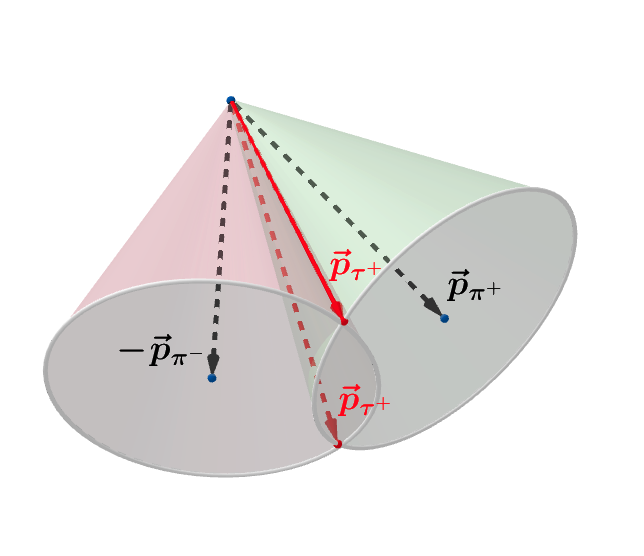}
  \caption{Geometric origin of the two-fold ambiguity in $\tau$ momentum reconstruction—intersection of two cones. The two red arrowed lines along the cone intersections indicate the possible $\tau$ momentum directions, while the black arrowed rays denote the momentum directions of the two pions.}
  \label{fig:placeholder}
\end{figure}

This geometric relationship is shown in Fig.\ref{fig:placeholder}, where the intersection of the two cones visually illustrates the origin of the double solution.

\subsection{Numerical Interpretation of the Two-fold Ambiguity Using SVD}
Next, we describe how to solve the above kinematic problem numerically. Since $\tau^+\tau^-$ are produced back-to-back in the whole physical process, we introduce a unit vector $\hat{u}$ representing the momentum direction of $\tau^+$ (or the opposite direction of the $\tau^-$ momentum). This vector satisfies the following equations:
\begin{equation}
\begin{aligned}
\hat{u}\cdot \hat{n}_{\pi^+} &= \cos\theta^+,\\
\hat{u}\cdot \hat{n}_{\pi^-} &= -\cos\theta^-,
\end{aligned}
\end{equation}
where $\hat{n}_{\pi^+}$ and $\hat{n}_{\pi^-}$ are the unit vectors of the $\pi^+$ and $\pi^-$ momenta, respectively, and $\cos\theta^+$ and $-\cos\theta^-$ are determined by the kinematic constraints given in Eq.~(\ref{eq:5}). The above system can be recast as a linear one:
\begin{equation}
A\hat{u}=b~,
\end{equation}
where
$$
A=
\begin{bmatrix}
\hat{n}_{\pi^+}^T \\
\hat{n}_{\pi^-}^T
\end{bmatrix}, \quad
b=
\begin{bmatrix}
\cos\theta^+ \\
-\cos\theta^-
\end{bmatrix}.
$$
Here $A$ is a $2\times 3$ matrix, which is not square and thus has no inverse. Consequently, conventional matrix inversion methods are not applicable. To solve this underdetermined system of equations, we introduce the singular value decomposition (SVD) method.

\subsubsection{Singular Value Decomposition (SVD) Framework}
By virtue of SVD, the matrix $A$ can be decomposed into three matrices~\cite{Sadek2012,Lieven:2006iut}:
\begin{equation}
A = U \Sigma V^T,
\end{equation}
where $U$ is a $2\times2$ square matrix, $\Sigma$ is a $2\times3$ matrix, and $V^T$ is a $3\times3$ matrix.
$U$ and $V$ are constructed from $A$ as follows:
for $AA^T$: obtained are the eigenvalues $\lambda_1$ and $\lambda_2$ with corresponding eigenvectors $u_1$ and $u_2$, which form the matrix $U$:
$$ U = \begin{bmatrix} u_1 & u_2 \end{bmatrix}~.$$
Similarly for $A^TA$, obtained are the eigenvalues $\lambda_1$, $\lambda_2$, $\lambda_3$ and their corresponding eigenvectors $v_1$, $v_2$, $v_3$, which form the matrix $V$:
$$
V = \begin{bmatrix} v_1 & v_2 & v_3 \end{bmatrix}~.
$$
Finally, construct $\Sigma$ by placing the singular values (square roots of the eigenvalues) on the diagonal:
$$
\Sigma =
\begin{pmatrix}
\sqrt{\lambda_1} & 0 & 0 \\
0 & \sqrt{\lambda_2} & 0
\end{pmatrix}~.
$$

To find a particular solution, we construct the pseudo-inverse $A^+$ of $A$, which can be expressed as:
$$
A^+ = V \Sigma^+ U^T,
$$
$A^+$ satisfies:
\begin{align}
& AA^+A = A, \\
& A^+AA^+ = A^+, \\
& (AA^+)^T = AA^+, \\
& (A^+A)^T = A^+A.
\end{align}
Then a particular solution $u_0$ is given by:
\begin{equation}
u_0 = A^+ b = V \Sigma^+ U^T b,
\end{equation}
where $$\Sigma^{+} = \begin{pmatrix} 
\frac{1}{\sqrt{\lambda_1}} & 0 \\ 
0 & \frac{1}{\sqrt{\lambda_2}} \\ 
0 & 0 
\end{pmatrix}.$$
When solving for the particular solution, the eigenvalue $\lambda_3$ is effectively ignored. To obtain the complete solution for the matrix $A$, a correction term $v_3$ must be added. Thus, the general solution of the above equation can be expressed as:
\begin{equation}
\hat{u} = u_{0} + \alpha v_3.
\end{equation}
Where $v_3$ is the null-space vector of $A$ (satisfying $A v_3 = 0$), and $\alpha$ is a coefficient to be determined. This degree of freedom originates from the kinematic incompleteness caused by the missing neutrino information.

\subsubsection{Unit Norm Constraint and the Number of Solutions}
For the general solution, we require it to be a unit vector (i.e., $|\hat{u}|^{2}=1$). This translates the general solution into a quadratic equation in $\alpha$:
\begin{equation}
\alpha^{2} |v_3|^{2} + 2\alpha (u_{0}\cdot v_3) + (|u_{0}|^{2} - 1) = 0,
\end{equation}
The discriminant of the above equation is:
\begin{equation}
\Delta = 4\left[(u_0\cdot v_3)^2 - |v_3|^2 (|u_0|^2 - 1)\right].
\label{eq:dis}
\end{equation}
If $\Delta > 0$, the equation has two real roots, corresponding to two feasible directions for $\hat{u}$. Geometrically, this corresponds to the intersection of the two cones along two lines, giving two possible $\tau$ directions.
If $\Delta = 0$, the equation has a single root, corresponding to a unique solution. Geometrically, this corresponds to the two cones being tangent.
If $\Delta < 0$, the equation has no real solution, indicating that the given kinematic input has no physical counterpart, i.e., the two cones are disjoint.

\subsubsection{Reconstruction Procedure}
\begin{enumerate}
    \item Construct the four-momenta from the visible particles.
    \item Construct the constraint matrix $A$ and vector $b$.
    \item For each event, solve using SVD to obtain the $\tau$ momentum.
\end{enumerate}

%%%%%%%%%%%%%%%%%%%%%%%%%%%%%%%%%%%%%%%%%%%%%%%%%%%%%%%%%%%
\section{Monte Carlo Simulation and Validation}\label{sec:3}
\subsection{Simulation Setup}
To verify the accuracy of the SVD method in reconstructing $\tau^+\tau^-$, we perform numerical experiments using Monte Carlo simulation software. For simplification, we only consider two important experimental parameters: the energy spread of the BEPC-II and the momentum resolution (for specific details, please refer to the appendix).

Energy simulation: the energy spread of the BEPC-II is $\sigma_E = 1.2 \, \text{MeV}$~\cite{BEPCII2003, Yuan2010}, in the rest frame of $\psi(2S)$, the nominal energy is \( W_0 \), the experimental energy is obtained by sampling  
\[
W = W_0 + \sigma_E \cdot \xi.
\]
Where \( \xi \) is a random number of standard normal distribution \( N(0,1) \).

Detector simulation: The momentum resolution of BEPC-II is $\delta / p = 0.5\%$~\cite{BESIII2004}. For simplicity, we simulate the detector resolution using
$$
p_{\text{meas}} = p_0 + \delta \cdot \zeta, 
$$
where $\zeta$ follows a standard normal distribution $N(0,1)$. A random $\zeta$ is drawn, multiplied by the momentum resolution $\delta$, and added to the nominal momentum $p_0$ to obtain the measured momentum $p_{\text{meas}}$. The direction is kept unchanged, and the components are recalculated using the new magnitude:

$$p_{i, \text{meas}} = p_{\text{meas}} \times \frac{p_i}{p_0}, \quad i = x, y, z.$$

We generate a total of 100,000 simulated events. For each event, we record:
\begin{enumerate}
    \item the true $\tau^+\tau^-$ (“MC truth-level” data);
    \item the momentum information of the visible decay products $\pi^+$ and $\pi^-$.
\end{enumerate}

\subsection{Reconstruction Efficiency Statistics}
Based on the kinematic information of $\pi^+$ and $\pi^-$, the $\tau$ momentum direction is reconstructed using the SVD method. Table~\ref{tab:reco_efficiency} summarises the reconstruction statistics for $1\times10^5$ simulated events. When beam energy spread and detector resolution are included, $89965$ events yield two distinct solutions, $10035$ yield no solution, and no event yields two identical solutions, giving a reconstruction efficiency of $89.97\%$.  

Without energy spread and detector resolution, the efficiency rises to $95.63\%$ ($95632$ successes, $4638$ failures). In all failed events, the kinematic parameters lie exactly near the boundary of inequality \eqref{eq:iq} (i.e., $\Omega >\theta_{\tau^+,\pi^+} + \theta_{\tau^+,-\pi^-}$ or $\Omega < |\theta_{\tau^+,\pi^+} - \theta_{\tau^+,-\pi^-}|$), and they should theoretically succeed; however, floating-point rounding errors cause discriminant \eqref{eq:dis} to become slightly negative. Among the $95632$ successful events, if we require the angle between the two reconstructed $\tau^+$ directions to be less than $10^\circ$, there are $4180$ events; these are also near-boundary events for which the discriminant accidentally evaluates to a positive value.
\begin{table}[htbp]
\centering
\caption{Reconstruction efficiency of the SVD method with and without beam energy spread and detector resolution}
\small
\setlength{\tabcolsep}{3pt}
\begin{tabular}{lccc}
\toprule
\textbf{Condition} & \textbf{No sol.} & \makecell{\textbf{Two distinct}\\ \textbf{solutions}} & \textbf{Eff.} \\
\midrule
\makecell{With spreads \\ and resolution} & 10035 & 89965 & 89.97\% \\
\makecell{Without spreads \\ and resolution} & 4638 & 95632 & 95.63\% \\
\bottomrule
\multicolumn{4}{l}{\footnotesize Total events: 100000 per condition. No two identical solutions.} \\
\end{tabular}
\label{tab:reco_efficiency}
\end{table}

\subsection{Validation of the Reconstruction Method Precision}

As can be seen from Table~\ref{tab:reco_efficiency}, all successfully reconstructed events exhibit the phenomenon of two-fold ambiguity. Under real experimental conditions, it is impossible to determine which of the two reconstructed solutions corresponds to the true physical process, owing to the lack of direct measurement of the $\tau$ lepton itself. However, in Monte Carlo (MC) simulation, the “MC truth-level" momentum information of the $\tau$ is fully recorded. This makes it possible to compare the two reconstructed solutions with the MC truth-level momentum, thereby enabling an unambiguous distinction between the true solution and the spurious solution. Accordingly, we construct three datasets: the MC truth-level $\tau$ momentum set, the reconstructed true-solution $\tau$ momentum set, and the reconstructed spurious-solution $\tau$ momentum set. The purpose of this section is to exploit this unique capability provided by the simulation to quantitatively assess the precision and reliability of the reconstruction method by examining the consistency between the reconstructed true solution and the MC truth-level results. The analysis of the spurious solution will be discussed separately in the following section.

In the validation process, we focus on the agreement of the following physical quantities:
\begin{itemize}
    \item The three Cartesian components of the $\tau$ momentum ($p_x, p_y, p_z$);
    \item The angular distribution between the direction of the reconstructed true $\tau$ and that of the  MC truth-level $\tau$;
    \item The scattering angle distribution in the $\tau$ production process.
\end{itemize}

\begin{figure}
  \centering
  \includegraphics[width=0.45\linewidth]{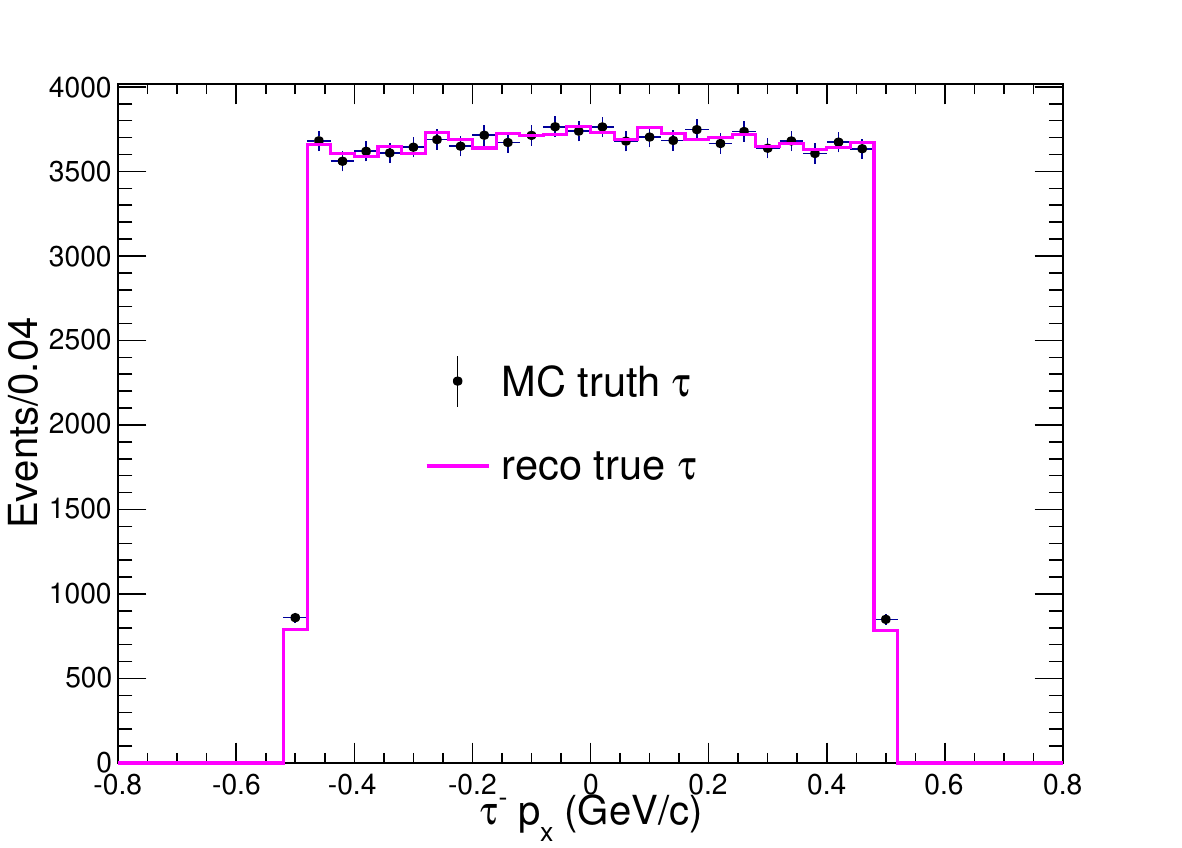}
  \qquad
  \includegraphics[width=0.45\linewidth]{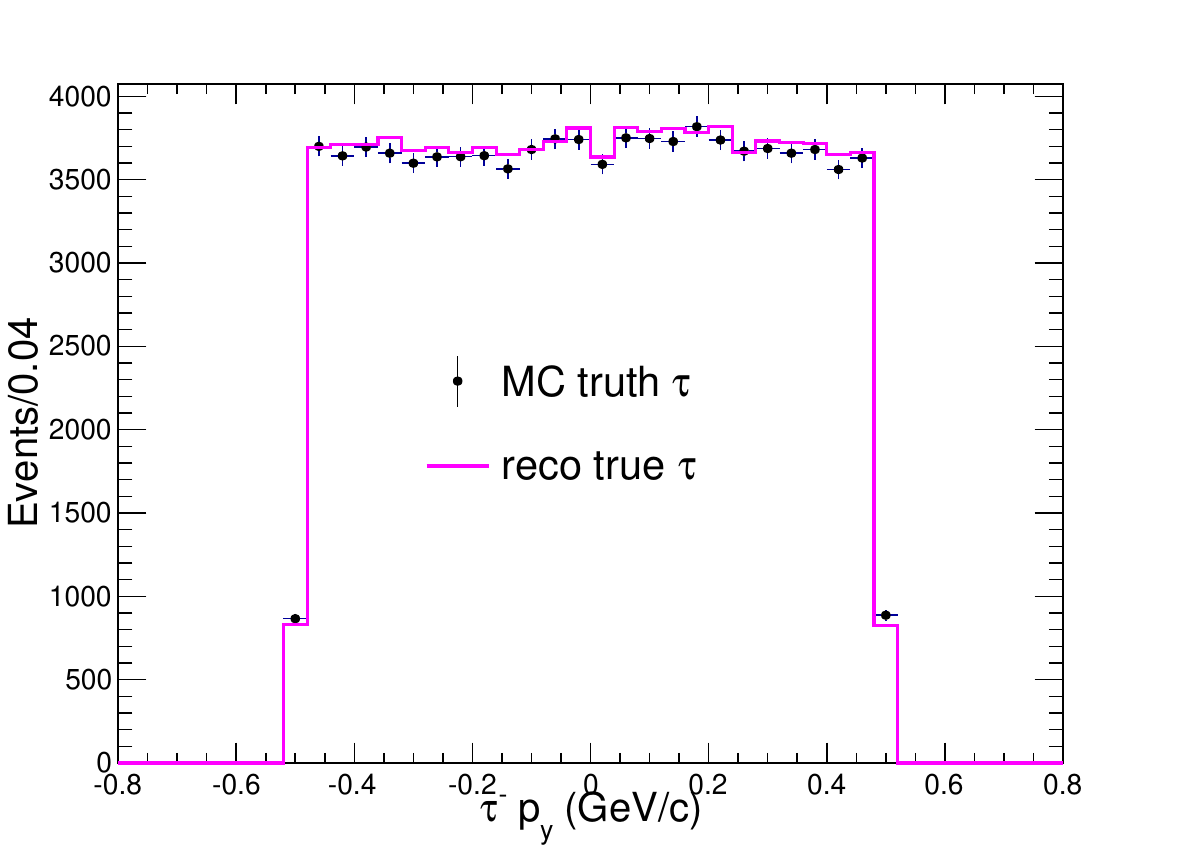}
   \qquad
   \includegraphics[width=0.45\linewidth]{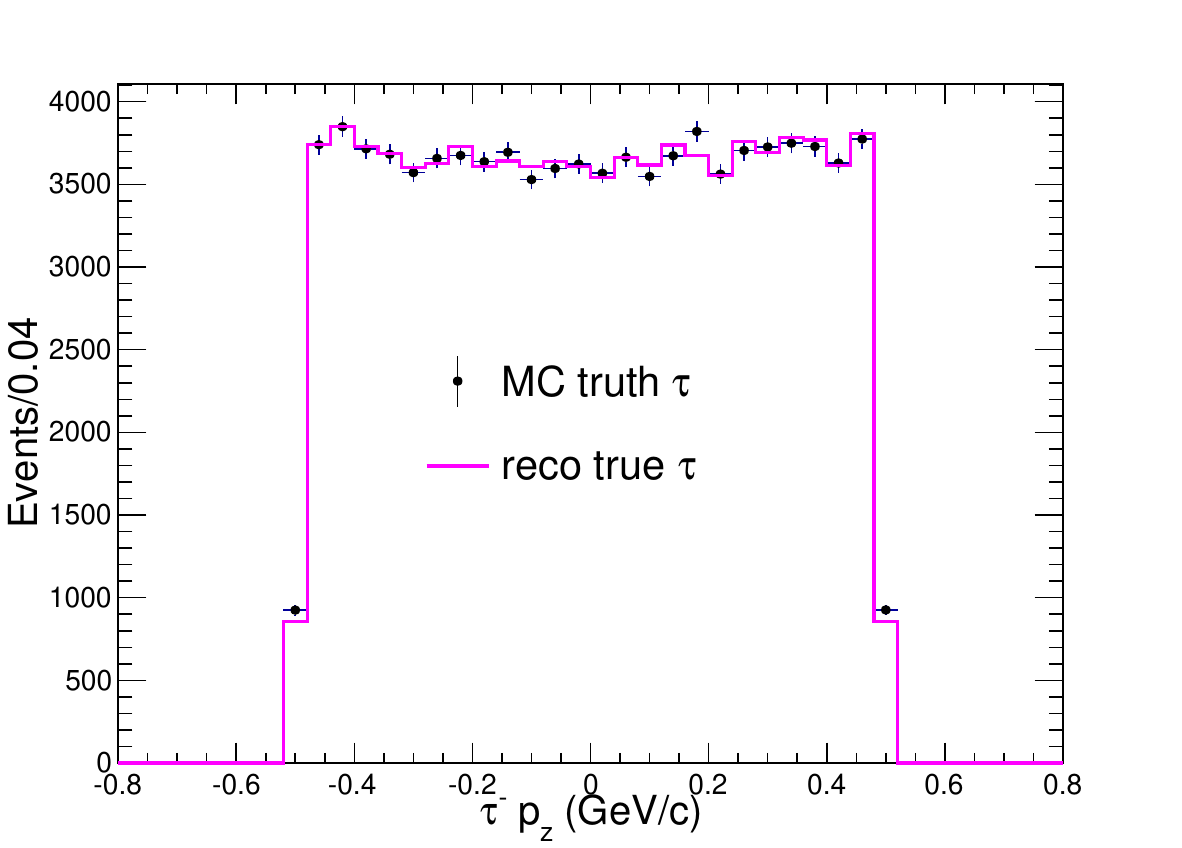}
    \qquad
    \includegraphics[width=0.45\linewidth]{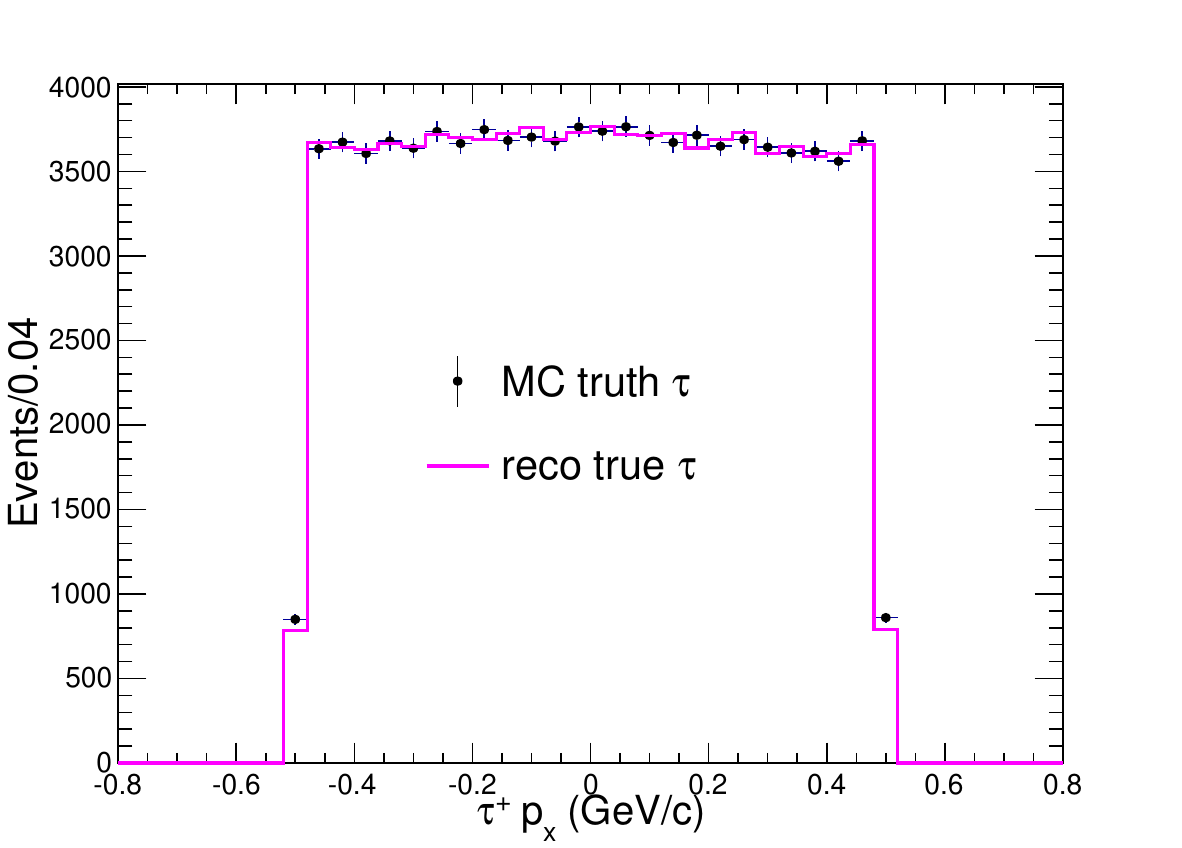}
     \qquad
     \includegraphics[width=0.45\linewidth]{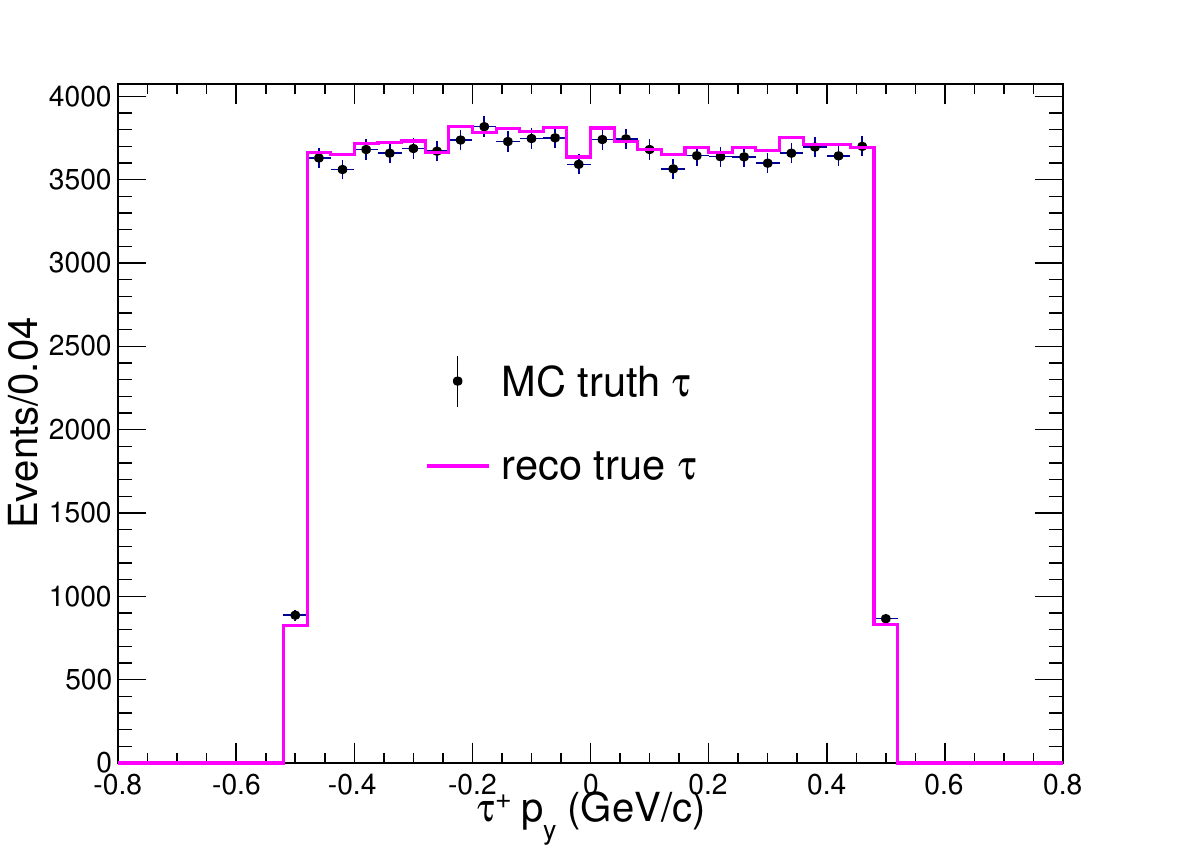} 
      \qquad
     \includegraphics[width=0.45\linewidth]{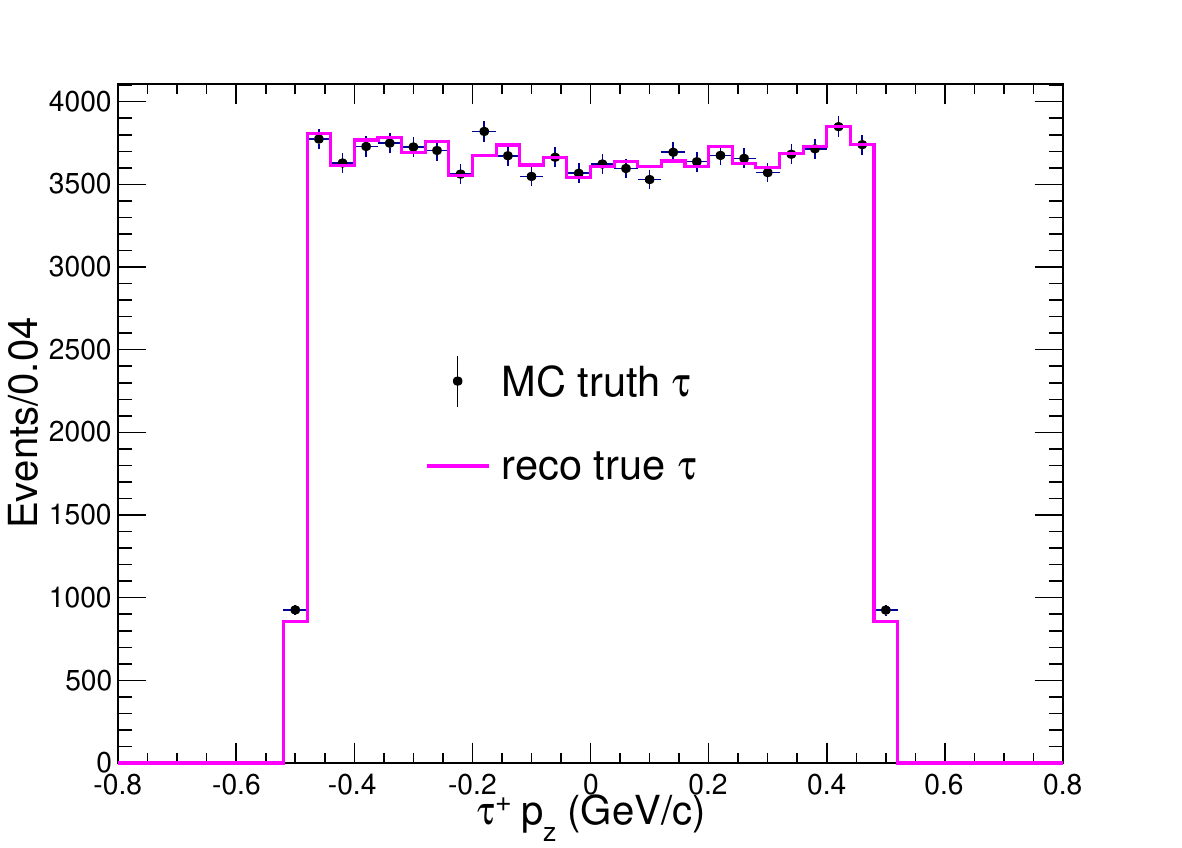}
      \centering
  \caption{Comparison of the Cartesian components of the $\tau$ momentum between the reconstructed true solution and the MC truth-level distribution. Black points with error bars represent the MC truth-level momentum distribution, while the red histogram corresponds to the reconstructed true solution.}
  \label{fig:mon}
\end{figure}

\begin{figure}
  \centering
  \includegraphics[width=0.45\linewidth]{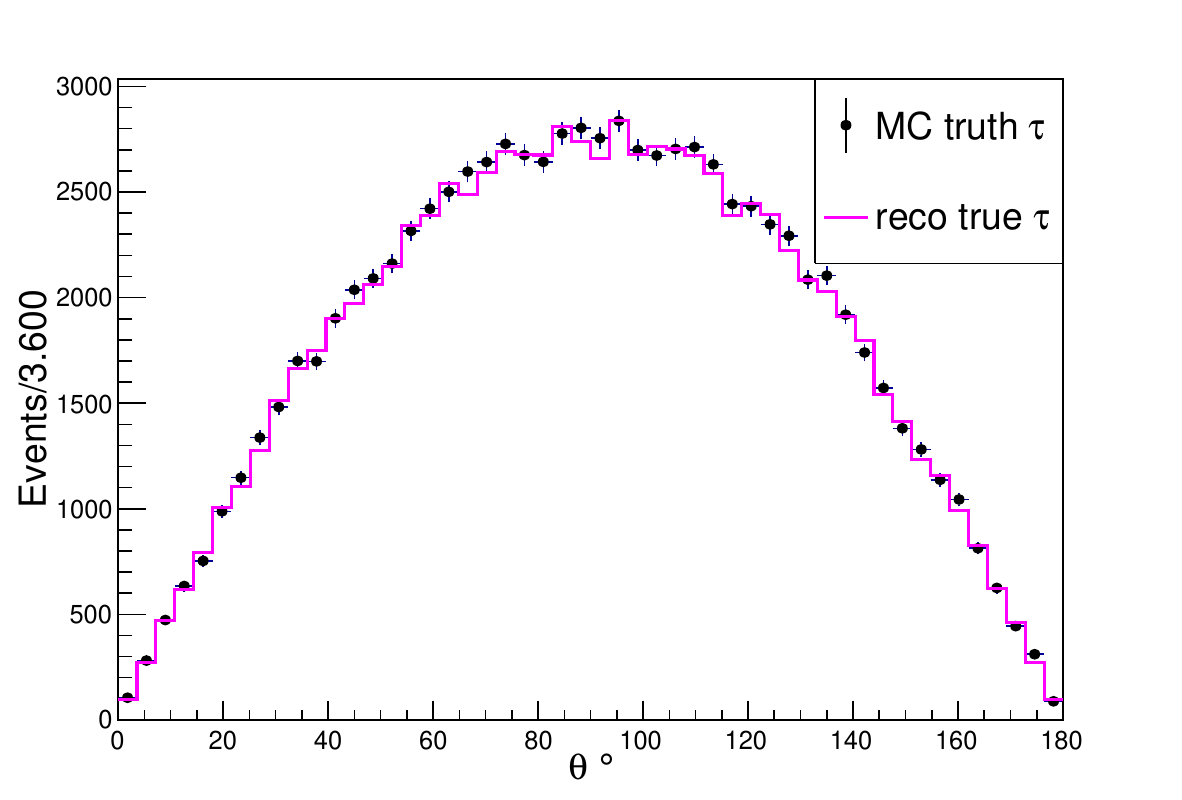}
  \qquad
  \includegraphics[width=0.45\linewidth]{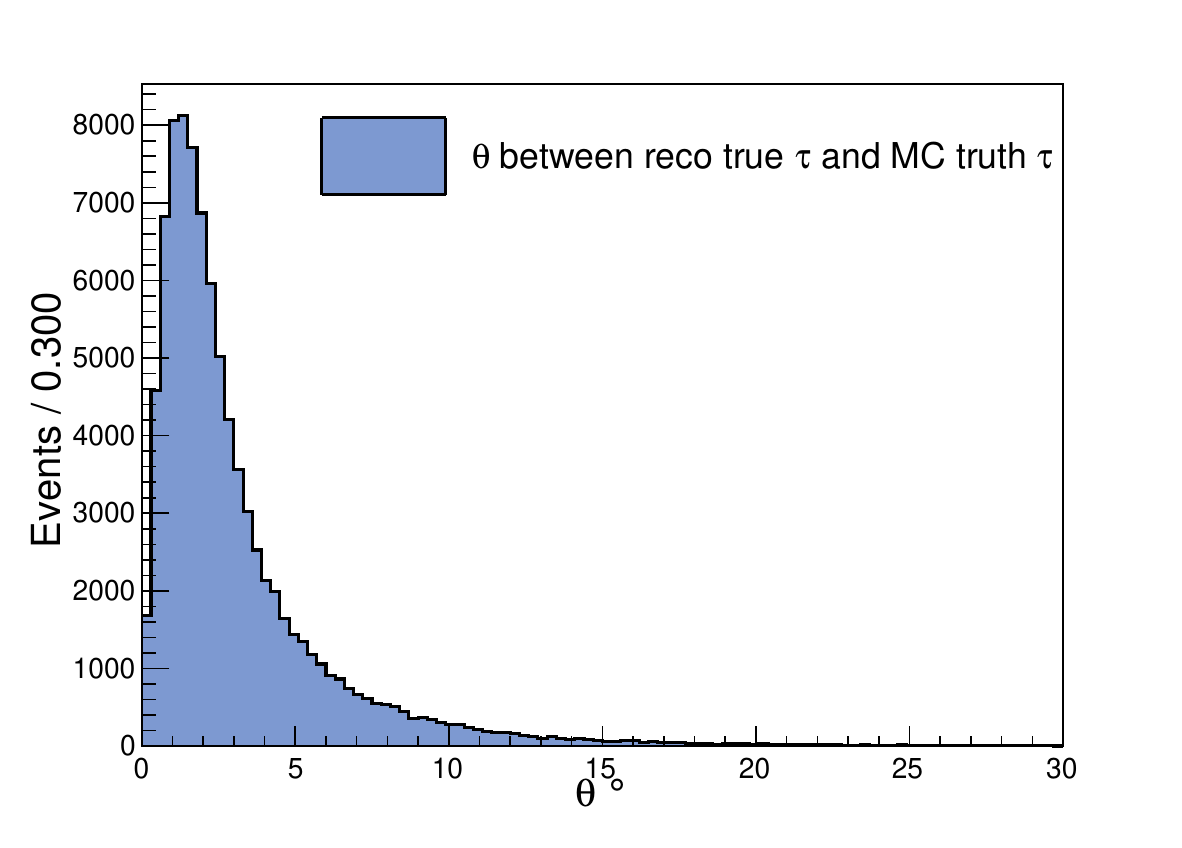}
  \caption{Comparison between the reconstructed true-solution $\tau$ momentum and the MC truth-level $\tau$ momentum. (left) Comparison of the scattering angle distributions in the $\tau$ production process: black points with error bars represent the MC truth-level distribution, while the red solid line corresponds to the reconstructed true-solution distribution. (right) Angular distribution between the reconstructed true $\tau$ momentum direction and the MC truth-level $\tau$ momentum direction.}
  \label{fig:theta}
\end{figure}

The corresponding comparison results are presented as histograms in Figs.~\ref{fig:mon} and ~\ref{fig:theta}, respectively, providing a direct visualization of the agreement between the reconstructed true solution and the MC truth-level values. Overall, the reconstructed true solution obtained via the SVD method exhibits a high degree of consistency with the  MC truth-level solution generated by the simulation, thereby validating the effectiveness of the reconstruction method.

\begin{enumerate}
    \item \textbf{Beam energy spread}: In the reconstruction process, the production energy of $\psi(2S)$ is assumed to be a fixed value of 3.686~GeV. However, in the actual simulation, the beam energy distribution  effects have been taken into account, leading to a certain broadening of the $\tau$ momentum itself, which introduces reconstruction biases.

    \item \textbf{Detector effects}: Due to the finite momentum resolution of the detector, the measured momentum deviates from the true value, resulting in a random distribution in the reconstructed $\tau$ momentum.
\end{enumerate}

The combined effect of these factors leads to a certain distribution in the angle between the reconstructed true $\tau$ direction and the MC truth-level $\tau$ direction. This feature is clearly shown in Fig.~\ref{fig:theta}.

%%%%%%%%%%%%%%%%%%%%%%%%%%%%%%%%%%%%%%%%%%%%%%%%%%%%%%%%%%%

\section{Application to Spin Entanglement Measurement}\label{sec:4}
\subsection{Spin Correlation Observables}
Using the $\tau$ lepton momenta reconstructed by the SVD method, we can further extract the spin correlations between the $\tau^+$ and $\tau^-$. For this system, the quantum state is fully described by a $4\times4$ density matrix $\rho$. Using the Fano-Bloch decomposition~\cite{Fano:1983zz}, we obtain:
{\small
\begin{equation}
\rho = \frac{1}{4}\Big( \mathbb{I}_2\otimes\mathbb{I}_2 + \sum_i B_i^+ \sigma_i\otimes\mathbb{I}_2 + \sum_j B_j^- \mathbb{I}_2\otimes\sigma_j + \sum_{i,j} C_{ij}\,\sigma_i\otimes\sigma_j \Big),
\end{equation}
}
where the summations run over \(i, j = 1, 2, 3\). Here \(\mathbb{I}_2\) is the $2\times2$ identity matrix, $\sigma_i$ are the Pauli matrices. The coefficients $B_i^+$ are the net polarizations of  $\tau^+$, $B_j^-$ are the net polarizations of  $\tau^-$, and $C_{ij}$ is the spin correlation matrix.

The concurrence is used to quantify the strength of entanglement. For a two-qubit system, it is defined as~\cite{Wootters:1997id}:
\begin{equation}
\mathcal{C}(\rho) = \max(0, \lambda_1 - \lambda_2 - \lambda_3 - \lambda_4),
\end{equation}
where $\lambda_i$ are the eigenvalues (sorted in descending order) of the auxiliary matrix $R$ .
\begin{equation}
R = \sqrt{\sqrt{\rho}\tilde{\rho}\sqrt{\rho}}, \quad \tilde{\rho} = (\sigma_2 \otimes \sigma_2)\rho^*(\sigma_2 \otimes \sigma_2).
\end{equation}
Separable states satisfy $\mathcal{C}(\rho) = 0$, while entangled states satisfy $0 < \mathcal{C}(\rho) \leq 1$. A larger $\mathcal{C}(\rho)$ indicates a higher degree of entanglement.

At the center-of-mass energy of BEPC-II, this expression simplifies to~\cite{Han:2025ewp,Cheng:2024rxi}:
\begin{equation}
\mathcal{C}(\bar{\rho}) = \frac{1}{2}(C_{11} + C_{33} - C_{22} - 1).
\label{C11}
\end{equation}
where $\bar{\rho}$ denotes the concurrence of the ensemble-averaged density matrix reconstructed from all events. 

Therefore, experimentally, one only needs to measure the corresponding $C$ matrix to compute the entanglement strength.

\subsection{Extraction of Fano Coefficients}
Here we adopt a kinematic method~\cite{Han:2025ewp} for extraction. At low energies $\sqrt{s} \ll m_Z$, the spin correlation matrix simplifies to~\cite{Ehataht:2023zzt}:
{\small
\begin{equation}
C_{ij} = 
\begin{pmatrix}
\frac{(2 - \beta^2) \sin^2 \theta}{2 - \beta^2 \sin^2 \theta} & 0 & \frac{\sqrt{1 - \beta^2} \sin 2\theta}{2 - \beta^2 \sin^2 \theta} \\[6pt]
0 & \frac{-\beta^2 \sin^2 \theta}{2 - \beta^2 \sin^2 \theta} & 0 \\[6pt]
\frac{\sqrt{1 - \beta^2} \sin 2\theta}{2 - \beta^2 \sin^2 \theta} & 0 & \frac{\beta^2 + (2 - \beta^2) \cos^2 \theta}{2 - \beta^2 \sin^2 \theta}
\end{pmatrix}_{\!ij},
\end{equation}
}
where $\beta$ is the velocity of the $\tau$ lepton in the $e^+e^-$ collision center-of-mass frame, and $\theta$ is the scattering angle of the $\tau$.       

Thus, to extract the Fano coefficients, one only needs the kinematic information of the $\tau$ leptons, after which the entanglement strength can be computed.

\subsection{Entanglement Computation}
\subsubsection{Treatment Of the Two-fold Ambiguity}
Among the Two-fold Ambiguity arising from the reconstruction process, besides the “true solution” corresponding to the actual physical process, the other solution also strictly satisfies the four-momentum conservation of the $\tau$ decay. This solution is kinematically self-consistent and can be regarded as another possible momentum configuration of the final-state particles.

Figure~\ref{fig:entanglement}  left shows a comparison of the angular distributions of the true and spurious $\tau$ solutions. Although the spurious solution does not match the true momentum on an event-by-event basis, its scattering angle distribution exhibits good agreement with the true distribution in a statistical sense. This property is significant for quantum entanglement measurements: whether using kinematic~\cite{Han:2025ewp, Cheng:2024rxi} or decay-angle-based~\cite{Afik2021} entanglement analysis methods, the key quantity depends on the statistical distribution of $\tau$ leptons.

Therefore, in large-sample experiments with sufficient statistics, retaining both solutions for subsequent analysis does not introduce systematic biases in entanglement measurements. For future experiments, if the collider is equipped with high-precision secondary vertex detectors, it will be possible to distinguish the double solutions on an event-by-event basis using vertex information or other kinematic correlations, further improving the accuracy of momentum reconstruction and the sensitivity of entanglement analysis.

\subsubsection{Entanglement Analysis with the Two-fold Ambiguity}
\begin{figure}
    \centering
    \includegraphics[width=0.45\linewidth]{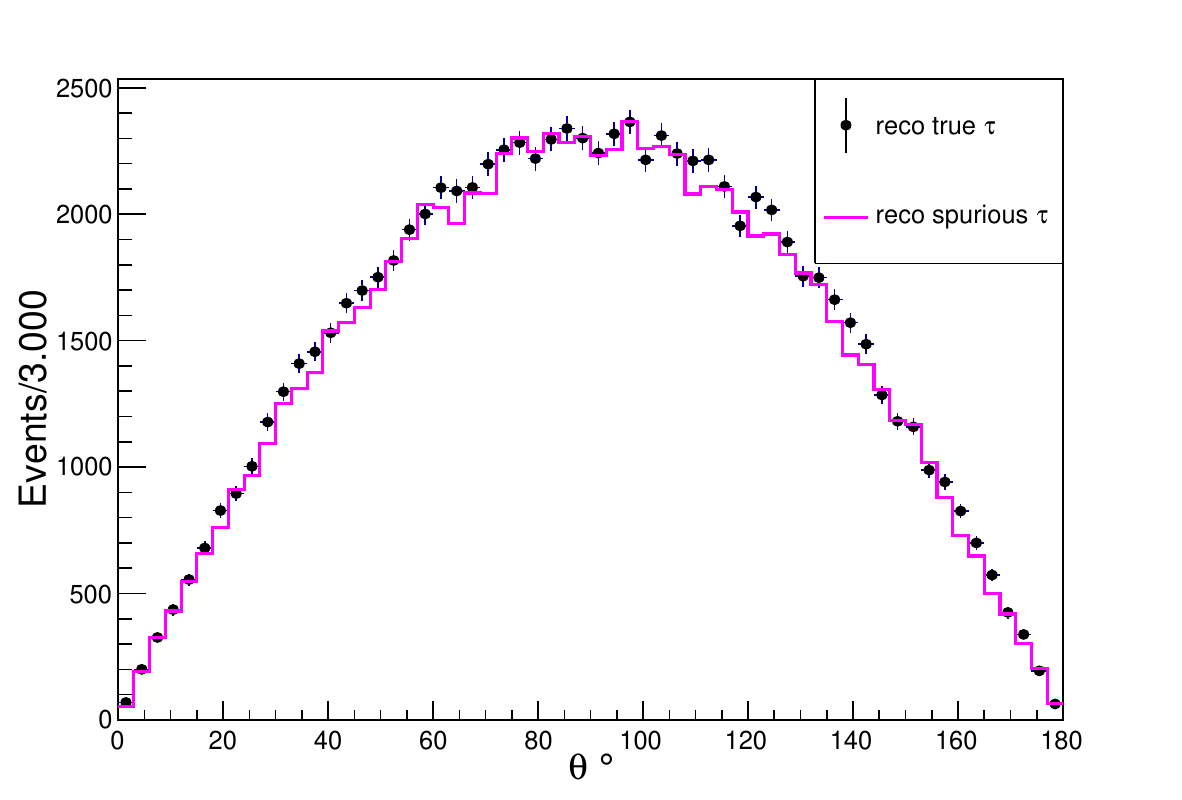} 
    \qquad
    \includegraphics[width=0.45\linewidth]{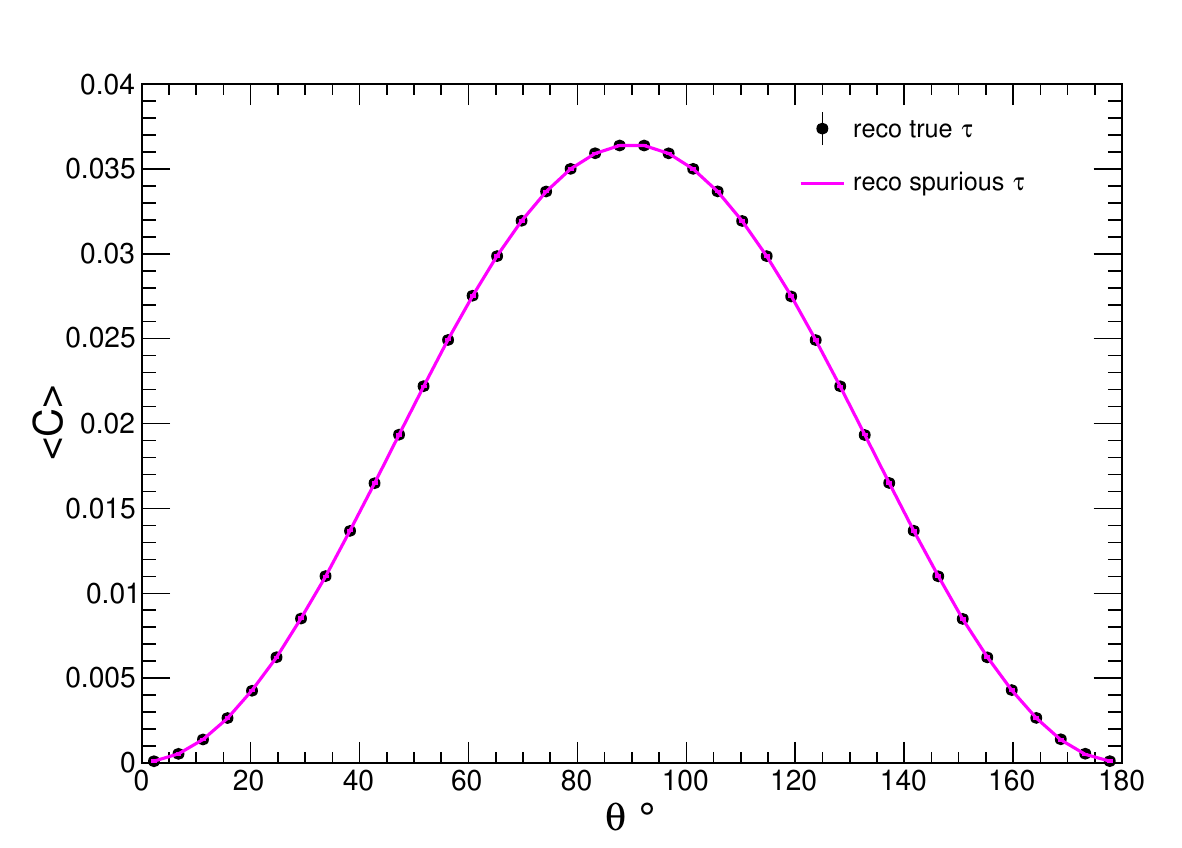}
    
    \caption{Comparison of the scattering angle distributions for the true and spurious $\tau^-$ solutions(left). Comparison of the concurrence distributions computed from the reconstructed true solution and the reconstructed spurious solution(right). }
    \label{fig:entanglement}
\end{figure}

The experimental procedure is as follows:
\begin{enumerate}
    \item Input the three-momenta of $\pi^{+}$ and $\pi^{-}$.
    \item Obtain the four-momenta of $\tau^{+}$ and $\tau^{-}$ (retaining both solutions).
    \item Compute the correlation matrix $C_{ij}$ and concurrence estimator $\mathcal{C}(\theta_i,\beta_i)$ of each event.
    \item Compute the concurrence as\footnote{In Eq.~\eqref{C11}, the concurrence $\mathcal{C}(\bar\rho)$ depends linearly on the spin-correlation components $C_{ij}$. Consequently, within the kinematic method, the ensemble-averaged concurrence can be obtained directly by averaging the event-by-event concurrence estimator $\mathcal{C}(\theta_i,\beta_i)$. This follows from the specific form of the density matrix considered in this work.
}
\begin{equation}
 \mathcal{C}(\bar{\rho})=\langle  \mathcal{C}(\theta,\beta)   \rangle=\frac{1}{N} \sum_{i=1}^{N} \mathcal{C}(\theta_i,\beta_i).
\end{equation}

\end{enumerate}
Following the above procedure, we obtain the entanglement results for the three datasets. Figure~\ref{fig:entanglement} presents a comparison of the concurrence distributions computed from the true-solution set and the spurious-solution set. As can be seen from the figure, the two distributions are in excellent agreement, corroborating the conclusion drawn in the previous section that the spurious solution exhibits high statistical consistency with the true solution in terms of physical observables.

Table~\ref{tab:C_compact_en} lists the comparison of the mean concurrence $\langle C \rangle$ (mean $\pm$ statistical uncertainty) for the true solution, spurious solution, and Monte Carlo truth in different angular bins. The results show that the spurious solution agrees well with the true solution. Therefore, even if the two solutions cannot be distinguished on an event-by-event basis, retaining both in the statistical analysis does not introduce bias in the entanglement measurement.
\begin{table*}[!t]
\centering
\caption{Comparison of the mean concurrence $\langle C \rangle$ (mean $\pm$ stat.) among the MC truth, true solution, and spurious solution in different $\theta$ bins}
\label{tab:C_compact_en}
\small
\setlength{\tabcolsep}{4pt}
\begin{tabular}{c c c c c c c}
\toprule
\multirow{2}{*}{$\theta$ bin (deg)} & \multicolumn{2}{c}{MC truth} & \multicolumn{2}{c}{True solution} & \multicolumn{2}{c}{Spurious solution} \\
\cmidrule(lr){2-3} \cmidrule(lr){4-5} \cmidrule(lr){6-7}
& Events & $\langle C \rangle \pm$ stat. & Events & $\langle C \rangle \pm$ stat. & Events & $\langle C \rangle \pm$ stat. \\
\midrule
0--15     & 1509 & $0.001202 \pm 0.000018$ & 1510 & $0.001194 \pm 0.000018$ & 1523 & $0.001204 \pm 0.000017$ \\
15--30    & 4523 & $0.005675 \pm 0.000028$ & 4546 & $0.005699 \pm 0.000028$ & 4457 & $0.005679 \pm 0.000028$ \\
30--45    & 7215 & $0.013516 \pm 0.000031$ & 7201 & $0.013562 \pm 0.000031$ & 7055 & $0.013538 \pm 0.000031$ \\
45--60    & 9202 & $0.022683 \pm 0.000030$ & 9196 & $0.022718 \pm 0.000028$ & 9249 & $0.022730 \pm 0.000028$ \\
60--75    & 10665& $0.030768 \pm 0.000023$ & 10711& $0.030878 \pm 0.000019$ & 10637& $0.030849 \pm 0.000019$ \\
75--90    & 11461& $0.035506 \pm 0.000017$ & 11373& $0.035620 \pm 0.000007$ & 11404& $0.035619 \pm 0.000007$ \\
90--105   & 11513& $0.035488 \pm 0.000018$ & 11580& $0.035626 \pm 0.000007$ & 11372& $0.035619 \pm 0.000007$ \\
105--120  & 10441& $0.030823 \pm 0.000023$ & 10445& $0.030888 \pm 0.000019$ & 10610& $0.030849 \pm 0.000019$ \\
120--135  & 9191 & $0.022707 \pm 0.000029$ & 9206 & $0.022749 \pm 0.000028$ & 9149 & $0.022741 \pm 0.000028$ \\
135--150  & 7144 & $0.013470 \pm 0.000031$ & 7074 & $0.013487 \pm 0.000031$ & 7233 & $0.013517 \pm 0.000031$ \\
150--165  & 4429 & $0.005644 \pm 0.000028$ & 4478 & $0.005664 \pm 0.000028$ & 4608 & $0.005692 \pm 0.000028$ \\
165--180  & 1564 & $0.001186 \pm 0.000018$ & 1537 & $0.001167 \pm 0.000017$ & 1560 & $0.001183 \pm 0.000018$ \\
\bottomrule
\end{tabular}
\end{table*}
%%%%%%%%%%%%%%%%%%%%%%%%%%%%%%%%%%%%%%%5

\section{Summary and Conclusion}
We analyzed the production and decay kinematics of $\tau^+\tau^-$ pairs in the rest frame of $\psi(2S)$, showing that the direction of each $\tau$ momentum is constrained to lie on a cone centered around the pion momentum direction. The geometric configuration of intersecting cones intuitively reveals the origin of the two-fold ambiguity.

To overcome the difficulty that conventional matrix methods cannot directly handle non-square systems, we introduced the singular value decomposition (SVD) method, converting the reconstruction equations into matrix form and obtaining the general solution via SVD. This method not only clearly reveals the numerical correspondence between the two-fold ambiguity and the null space structure of the matrix, but also uses a discriminant to reflect the relationship between the number of solutions and the kinematic configuration, providing a stable and reliable mathematical framework for $\tau$ lepton pair momentum reconstruction.

The SVD method was comprehensively validated using Monte Carlo simulated samples. The results show that the method can effectively reconstruct the momentum direction of the $\tau$ lepton, and the true solution of the two-fold ambiguity is in good agreement with the simulated truth in both momentum and scattering angle distributions. The observed finite deviations mainly originate from physical effects such as beam energy spread and detector resolution, consistent with expectations.

Although the SVD reconstruction suffers from the two-fold ambiguity, the spurious solution exhibits a high degree of consistency with the true solution in key distributions (e.g., the angular distribution of decay products in the $\tau$ rest frame). Consequently, in the calculation of spin entanglement quantifiers such as concurrence $\mathcal{C}(\bar\rho)$, the results obtained from the true and spurious solutions are statistically highly consistent. This finding has important experimental implications: even if current collider experiments are not equipped with high-precision secondary vertex detectors (and thus cannot directly distinguish $\tau$ decay vertices to reject the spurious solution), researchers do not need to select or discard either of the two-fold ambiguity solutions output by the SVD method; both can be retained and directly used for spin correlation measurements. This greatly reduces the requirements on detector hardware and simplifies the data analysis workflow. In large-sample analyses with sufficient statistics, retaining the two-fold ambiguity does not introduce systematic bias into the quantum entanglement measurement. Thus, the SVD-based reconstruction method established in this work not only provides a numerically stable and automatable approach for reconstructing $\tau$ lepton momenta, but more importantly implies that existing data from currently running collider experiments such as BESIII and Belle II can be readily used to study spin entanglement in $\tau$ lepton pairs, without waiting for future detector upgrades. Our work solves a key technical challenge in $\tau$ momentum reconstruction and opens a practical, immediate path toward quantum entanglement experiments in the lepton sector.

\acknowledgments
Xiang Zhou thanks Jintao Chen for useful discussions. Xiang Zhou, Xia Wan, and Youkai Wang are supported by the National Science Foundation of China under Grant No. 11947416.

Jianyong Zhang and Xiaohu Mo are supported in part by the National Key R\&D Program of China under Grant No. 2023YFA1606000.
%%%%%%%%%%%%%%%%%%%%%%%%%%%%%%%%%%%%%%%5
\appendix
\section{Details of the Monte Carlo Simulation}
\label{sec:apx}

The Monte Carlo (MC) simulation for the process \(e^+e^- \to \psi(2S)/\gamma^* \to \tau^+\tau^- \to \pi^+\bar{\nu}_\tau \pi^-\nu_\tau\) is performed in five steps, including the coherent interference between the \(\psi(2S)\) resonance and the QED continuum.

\subsection{Total Cross Section and Interference}

The Born cross section for \(e^+e^- \to \tau^+\tau^-\) is described by the coherent sum of a continuum amplitude (virtual photon exchange) and a resonance amplitude (\(\psi(2S)\)):
\[
\sigma(s) = \frac{s}{12\pi} \left| \frac{e^2}{s} + \frac{g_{V,\tau}g_{V,e}}{s - m_V^2 + i m_V \Gamma_V} \right|^2 
           \beta \left(1 + \frac{2m_\tau^2}{s}\right),
\]
where \(s\) is the squared center-of-mass (c.m.) energy, \(m_V\) and \(\Gamma_V\) ~\cite{Navas2024} are the mass and total width of the \(\psi(2S)\), \(g_{V,\tau}g_{V,e}=3.1\times10^{-5}\) is the product of coupling constants determined from the BESIII measurement of the peak cross section (52 nb)~\cite{Achasov2019}, and \(\beta=\sqrt{1-4m_\tau^2/s}\). The fine-structure constant is denoted by \(\alpha=e^2/4\pi\).

\subsection{Sampling of the c.m. Energy with Beam Spread}

The beam energy spread of BEPC-II is approximately Gaussian with width \(\sigma_E= 1.2\ \mathrm{MeV}\). For each event the actual c.m. energy \(W\) is sampled as

\[
W = W_0 + \sigma_E \cdot \xi,\qquad \xi \sim \mathcal{N}(0,1),
\]
where \(W_0 = 3.686\ \mathrm{GeV}\) is the nominal energy. If \(W < 2m_\tau\) the event is discarded and resampled. To account for the energy dependence of the cross section, the event is accepted with probability \(\sigma(W)/\sigma_{\max}\), where \(\sigma_{\max}\) is the maximum of \(\sigma(W)\) within \(W_0\pm5\Delta\). This acceptance–rejection procedure ensures that the final energy distribution is proportional to \(\sigma(W)\) convoluted with the Gaussian spread.

\subsection{Angular Distribution and Polarization in \(\tau^+\tau^-\) Production}

In the c.m. frame, the direction of the \(\tau^-\) is characterized by the polar angle \(\theta\) relative to the electron beam axis. The normalized angular distribution is given by the QED formula (including the \(\tau\) mass effect):

\[
\frac{1}{\sigma}\frac{d\sigma}{d\cos\theta} \propto 1 + \cos^2\theta + \frac{4m_\tau^2}{s}\sin^2\theta.
\]

The azimuthal angle \(\phi\) is uniformly distributed in \([0,2\pi)\). The value of \(\cos\theta\) is sampled using the acceptance–rejection method with a uniform proposal distribution. From the sampled angles and the c.m. energy, the four-momenta of \(\tau^-\) and \(\tau^+\) are constructed (they are back-to-back). The magnitude of the \(\tau\) momentum is \(p_\tau = \frac{W}{2}\beta\).

The longitudinal polarization of the \(\tau^-\) along its flight direction is

\[
P_{\tau^-} = -\frac{\sin^2\theta}{1+\cos^2\theta + \frac{4m_\tau^2}{s}\sin^2\theta},
\]

and the polarization of the \(\tau^+\) is opposite: \(\vec{P}_{\tau^+} = -\vec{P}_{\tau^-}\).

\subsection{Decay of \(\tau\) Leptons to pions}

Only the decays \(\tau^- \to \pi^-\nu_\tau\) and \(\tau^+ \to \pi^+\bar{\nu}_\tau\) are simulated. In the rest frame of the \(\tau\), the decay is two-body; the pion energy is fixed:

\[
E_\pi^{\text{rest}} = \frac{m_\tau^2 + m_\pi^2}{2m_\tau},\qquad 
p_\pi^{\text{rest}} = \sqrt{(E_\pi^{\text{rest}})^2 - m_\pi^2}.
\]

The angular distribution of the pion with respect to the \(\tau\) polarization direction is linear due to the \(V-A\) interaction:

\[
\frac{d\Gamma}{d\cos\theta_\pi} \propto 1 + \alpha_\pi P_\tau \cos\theta_\pi,
\]

Where \(\alpha_\pi = +1\) for \(\tau^- \to \pi^-\nu_\tau\) and \(\alpha_\pi = -1\) for \(\tau^+ \to \pi^+\bar{\nu}_\tau\). The angle \(\theta_\pi\) is sampled using the acceptance–rejection method; the azimuthal angle \(\phi_\pi\) is uniform in \([0,2\pi)\).

\subsection{Lorentz Transformation to the Laboratory Frame}

The four-momentum of the pion in the laboratory frame is obtained by boosting the rest-frame four-momentum with the \(\tau\) velocity. If the \(\tau\) has four-momentum \((E_\tau,\vec{p}_\tau)\) in the lab frame, then the boost velocity is \(\vec{\beta} = \vec{p}_\tau/E_\tau\) and \(\gamma = E_\tau/m_\tau\). The lab-frame pion four-momentum is

\[
\begin{aligned}
E_\pi^{\text{lab}} &= \gamma \left(E_\pi^{\text{rest}} + \vec{\beta}\cdot\vec{p}_\pi^{\,\text{rest}}\right),\\
\vec{p}_\pi^{\,\text{lab}} &= \vec{p}_\pi^{\,\text{rest}} + \frac{\gamma-1}{\beta^2}\left(\vec{\beta}\cdot\vec{p}_\pi^{\,\text{rest}}\right)\vec{\beta} + \gamma E_\pi^{\text{rest}}\vec{\beta}.
\end{aligned}
\]

This transformation is applied separately for the \(\tau^-\) and \(\tau^+\), yielding the laboratory momenta of the \(\pi^-\) and \(\pi^+\). The final output for each event consists of the three-momenta of \(\tau^-\), \(\pi^-\), \(\tau^+\), and \(\pi^+\) (12 numbers) in units of \(\mathrm{GeV}/c\).

\bibliographystyle{apsrev4-1}
\bibliography{refs}
\end{document}